\documentclass[runningheads,a4paper]{llncs}
\usepackage{graphicx}
\usepackage{subcaption}
\usepackage{hyperref}
\begin{document}

\mainmatter  

\title{LLMs as Compiler for Arabic Programming Language}
\titlerunning{LLMs as Compilers for Arabic Programming Language}

\author{Serry Sibaee, Omar Najar, Lahouri Ghouti, Anis Koubaa}
\authorrunning{Serry Sibaee, Omar Najar, lghouti, akoubaa}

\institute{RIOTU labs, Prince Sultan University, KSA\\
\email{\{ssibaee, onajar,lghouti, akoubaa\}@psu.edu.sa}}

\maketitle  

\begin{abstract}
In this paper we introduce APL (Arabic Programming Language) that uses Large language models (LLM) as semi-compiler to covert Arabic text code to python code then run the code. Designing a full pipeline from the structure of the APL text then a prompt (using prompt engineering) then running the prodcued python code using PyRunner. This project has a three parts first python library \href{https://github.com/riotu-lab/apl-compiler}{(GitHub)}, a playground with simple interface and this research paper.

\end{abstract}

\section{Introduction}

Programming languages serve as crucial tools for communicating with computers, facilitating various tasks and applications. Over the past few decades, the landscape of programming languages has evolved significantly, with numerous languages emerging, predominantly in English for example python, java, php and C++.

In [5][6] the evolution and diversity of programming languages are explored through a historical lens. Beginning with Zuse's Plankalkül, the narrative progresses through seminal moments in programming language development, such as the introduction of pseudocodes, the emergence of Fortran for the IBM 704, and the advent of functional programming with LISP. Key languages like ALGOL 60, COBOL, and BASIC are examined for their contributions to business computing and timesharing systems. The text delves into the expansive capabilities of languages like PL/I, APL, and SNOBOL, which catered to diverse programming needs. Additionally, it highlights significant advancements such as the introduction of SIMULA 67, ALGOL 68's orthogonal design principles, and the inception of logic-based programming with Prolog. Noteworthy efforts in language design, including Ada's comprehensive approach and Smalltalk's pioneering object-oriented paradigm, are discussed alongside popular languages like C++ and Java. The chapter also explores scripting languages, markup/programming hybrid languages, and the prominence of C sharp as a flagship language within the .NET ecosystem. And as we can notice that all of the previous programming langauges are in English, and this shortage made developers enthusiatec about building their own programming languages with their mother-tounge languages, for example "Chinese BASIC" programming language in chinese, "Superlogo" in Dutch, "LSE (Language Symbolique d'Enseignement)" in french and "Teuton" in German [7]. 

Then we have several Arabic-based programming languages have been introduced over the years. One of the earliest endeavors in this domain was made by Al Alamiah company in 1987, with the creation of Arabic Sakhr Basic [9]. This language, an adaptation of BASIC, featured Arabic keywords and expressions tailored for MSX home computers. ARLOGO [10] emerged as another Arabic programming language, based on UCB Logo, primarily aimed at educational purposes. While ARLOGO is open-source and currently compatible only with Microsoft Windows, ARABLAN [11] was developed in 1995 specifically for teaching programming to school children in Arab countries. Al-Risalah [12] followed, offering an Arabic object-oriented programming experience inspired by languages like Pascal, C++, and Eiffel, with a focus on educational use. More recently, additional Arabic programming languages have been developed, including AMORIA[13], Ebda3[14], Jeem[15], Loughaty[16], Qlb[17], and Kalimat [18], And recently in 2019 we saw "Phoenix" The Arabic Object-Oriented Programming Language[19] These languages collectively contribute to the growing landscape of programming education and practice in Arabic-speaking communities.
Talking about programming langauges is not complete without talking about The history of programming language compilers because it is marked by significant milestones and components. The concept of a compiler was introduced by Grace Hopper in the early 1950s, with the A-0 System. The first high-level programming language to be developed was Fortran, created by John Backus and his team at IBM in the mid-1950s, which aimed to simplify the programming process and improve code efficiency [20] The evolution of compilers is closely tied to the development of programming languages and the diversification of computing needs. Key components of a compiler include lexical analysis, syntax analysis, semantic analysis, optimization, and code generation. Lexical analysis breaks down the source code into tokens, syntax analysis constructs a parse tree based on the language's grammar, semantic analysis checks for semantic consistency, optimization improves the performance of the generated code, and code generation translates the optimized intermediate representation into the target language [21]. The history of compilers is also intertwined with the evolution of software engineering practices and the quest for more reliable and efficient programming paradigms [22].

Then we have the "Elephant in the room" the Pre-trained language models (PLMs) that revolutionized various natural language processing (NLP) tasks by capturing rich linguistic representations. That could be summarized focusing on their architectural differences and key families.

Encoder-only models, like BERT and its variants, excel at language understanding tasks like text classification. They process input text through an encoder network, learning contextual representations for downstream tasks. Decoder-only models, exemplified by GPT-1 and GPT-2, specialize in text generation. They employ a decoder network to iteratively produce text sequences conditioned on a prompt or previous token.

Encoder-decoder models, like T5 and BART, combine both capabilities. They utilize an encoder-decoder architecture, enabling them to handle understanding and generation tasks seamlessly. This unified approach makes them versatile for various NLP applications.

Beyond these established architectures, new families are emerging with unique characteristics. The LLaMA family emphasizes open-sourcing, fostering community-driven development of better LLMs. PaLM focuses on scaling, achieving state-of-the-art performance on numerous benchmarks. Other families like FLAN, Gopher, and ERNIE 3.0 showcase diverse design choices and research directions.

This overview highlights the multifaceted nature of PLMs, offering researchers and practitioners a range of options for tackling different NLP challenges. As the field continues to evolve, further exploration of novel architectures and training methods is expected to push the boundaries of language understanding and generation capabilities [23]. 

After this introduction we can see that there are some connections between programming languages compilers and LLMs in the general (abstract) where both has a hidden structure to solve (work with) problems so the idea to use LLMs as a semi programming compiler that translate Arabic structured text (APL programming language) to a Python code that could be run. We will discuss the architecture of the compiler where we will see the changes of the writing style of the APL, then we discuss the capabilities of the language. After that the prompt engineering of the LLMs that will be used to change the APL code to a Python code. Then Introducing a new API and interface of the APL that could be used to help who wants to use the compiler in their applications. 

\section{Related work}
In the dynamic realm of contemporary programming, the advent of Large Language Models (LLMs) parallels historical shifts in programming paradigms. Analogous to the evolution from assembly language to high-level languages facilitated by optimizing compilers, LLMs signify a transformative leap, positioning themselves not merely as advanced autocompletion tools but as compilers. This conceptual shift is underscored by the absence of direct exploration in the literature regarding the utilization of LLMs as programming compilers.

Historically, programming proficiency resided in mastering assembly language, a skill rendered obsolete by optimizing compilers translating high-level languages into efficient machine code. A parallel transition is underway today, elevating the act of programming to a higher level of abstraction. LLMs, conceived as compilers, signify this evolution by translating natural language prompts into high-level programming languages, redefining the coding landscape.

Empirical evidence, exemplified by Copilot, accentuates this transformative potential. Copilot demonstrated efficacy in solving introductory programming problems, showcasing an ability to generate accurate solutions from problem descriptions alone. In instances of inaccuracies, developers prefer prompt manipulation over conventional debugging, reflecting a paradigm shift in code creation methodologies.

The academic discourse delves into this trans-formative trajectory, with LLMCompiler [1] for parallel function calling and the innovative Synthesize, Execute, Debug (SED) approach [3] exemplifying engagements with evolving programming paradigms. Furthermore, critical insights from online articles [2] contribute nuanced perspectives, collectively consolidating the conceptualization of LLMs as compilers in the intricate landscape of modern programming methodologies. The absence of direct exploration in current literature underscores the nascent nature of employing LLMs as programming compilers, paving the way for future inquiries into this transformative domain.

Also there is an opposite approach of make a programming language that ease the use of LLM in python code called 'LMQL' "LMQL offers a novel way of interweaving traditional programming with the ability to call LLMs in your code."[4] And a side note that no one has tried explicitly to use LLMs as a compiler to run a new programming language but the general research shows to use LLMs to automate programming or help programmers to program fast with multiple way auto complete the written code (copilot) or solving the problems by asking the chat Bot to debug and suggest solutions. And the near idea that is similar to ours is "PseuToPy" [24] which is an explicit compiler (did not use LLMs but the idea is the same) that change chinese code which follows the structre of python code then run it. Also a new Approch that shows a reverse way of using LLMs as a decompilers where the LLM was Decompile-Eval dataset improve decompilation. [25]

\section{Methodology}
the methodology has three main parts the paper will explain each one of them separately 
The methodology unfolds in two key dimensions: Compiler Architecture and Prompt Engineering. Compiler Architecture investigates the integral components—LLM Planner, Task Fetching Unit, and Executor—providing a foundational understanding of LLM Compiler's structure. In parallel, Prompt Engineering explores the intricacies of transforming natural language prompts into effective instructions for the LLM, ensuring a holistic methodology that integrates both architectural foundations and linguistic nuances.

\subsection{Architecture of the compiler}
This section will discuss the structure of the programming language where how the input is represented and how the included operations (loops and functions).
\begin{figure}[htb]
    \centering
        \centering
        \includegraphics[width=\textwidth]{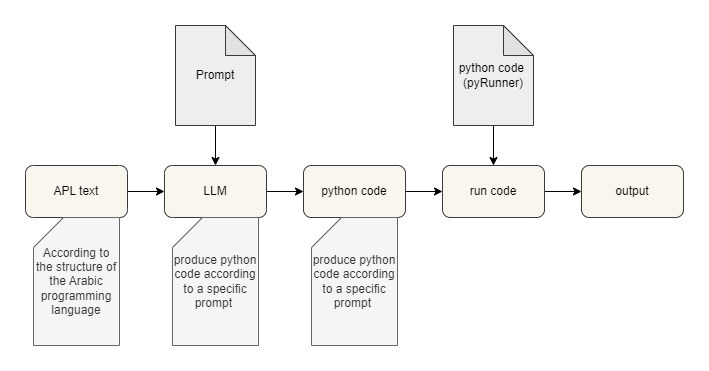}
        \caption{The architecture of APL compiler}
        \label{fig:1}
\end{figure}

\subsubsection{Structure of the input code}
The APL code is written completely in Arabic script from right to left. And for Loops (while for example) we chose not to use indentation of python but the curly brackets because it is easier for the programmer and also easier for the LLM to compiler because copying and pasting the indented text could mess the code. Also the final condition is to make sure the programmer write the code line by line and do not use the ';' to make it clearer and easier for the reader and for the LLM
\begin{enumerate}
    \item Arabic Script
    \item Curly Brackets Usage
    \item Line-by-Line Operations
\end{enumerate}
The code input follows an Arabic script, introducing a distinctive structure influenced by a shift to the use of curly brackets "\{\}" instead of indentation, departing from the traditional indentation in Python. Each operation is distinctly presented on its own line, eliminating the need for semicolons (;) in multi-line statements.
\subsubsection{Programming Language Components}
For simplicity we did not include all of python properties but just these six generic ones to make it simple and also for the fragility of LLMs in the chain of combined prompts. So these are the capabilities of the APL.
\begin{enumerate}
    \item Variables
    \item Operations
    \item File Reading (Limited to TXT Files)
    \item If Conditions
    \item Loops
    \item Functions
\end{enumerate}

\subsubsection{Arabic Text Conventions}
This section details the conventions employed for Arabic text within the programming language, including variable names, operations, conditions, and loops, as well as the syntax for functions and file reading.
\begin{figure}
    \centering
    \includegraphics[width=0.75\linewidth]{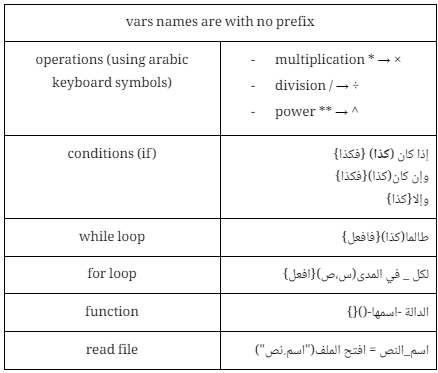}
    \caption{Functions in Arabic words in APL}
    \label{fig:enter-label}
\end{figure}
\subsection{Prompt Engineering: Dealing with the Text}
\subsubsection{Structure of the Prompt}
In the realm of prompt engineering, the task involves translating Arabic text into executable Python code. The process includes renaming variables to English for operational clarity and substituting Arabic functions with their Python equivalents (e.g., replacing  with print). The overarching goal is to ensure the prompt results in valid and runnable Python code, facilitating seamless integration within the programming environment.

\subsubsection{Prompt Structure Guidelines: Dealing with Keywords}
In the process of crafting prompts for translation into Python code, it is essential to handle specific Arabic terms and translate them into their corresponding Python keywords. The following table outlines the mapping of relevant keywords, ensuring their seamless integration into Python code while preserving Arabic texts used within strings.

\subsubsection{The used prompt}
The used prompt in the GPT-4 is: "This GPT specializes in translating Arabic algorithmic instructions into Python code. It accurately replaces Arabic keywords with their Python equivalents, ensures variables are named in English for runnable code, and checks for syntax correctness. If a syntax error is detected, it provides a specific error message to help users correct the issue. The output code is provided as plain text, not in Python markdown, to ensure compatibility with various text editors and environments."
\begin{figure}
    \centering
    \includegraphics[width=0.3\linewidth]{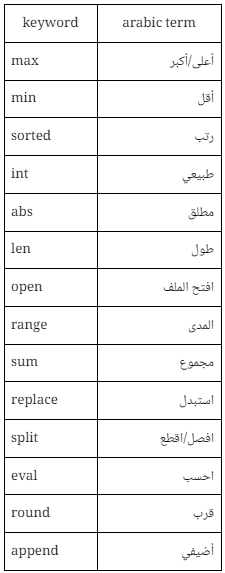}
    \caption{Keywords from the original language to Arabic APL}
    \label{fig:enter-label}
\end{figure}
\section{API and interface}

In this work, we conceptualize the API similarly to a traditional compiler. But rather than translating a high-level language into machine code, our API acts as a pipe, turning Arabic input into Python code. This method makes use of Python's adaptability and widespread use in AI applications and research.

\subsection{The Structure of the API}

The API is designed to be seamlessly integrated into existing Python environments. This
integration is achieved through a dedicated Python library, which is hosted on GitHub for accessibility and collaborative development. The library encapsulates the API's functionality into a single, user-friendly class. The main component of our Python library is the Compiler class. This class is initialized with an API key, enabling secure and authenticated access to the API's functionalities. The syntax is straightforward:

\subsection{Sample of the Interface}

To demonstrate the practical application and user interaction with our API, we introduce a sample interface designed for simplicity and effectiveness.

\textbf{Design of the Interface:} The interface features a dual-pane layout. The right pane serves as the input area where users can write or paste Arabic code. This design choice caters to the natural right-to-left flow of Arabic script, ensuring a familiar and comfortable user experience.

\textbf{Compatibility of Arabic Text with Output Display:} The output, or translated Python code, is shown in the left pane. A special focus is on making sure that Arabic text is compatible. This includes proper rendering of Arabic characters and script peculiarities in the output display in addition to left-to-right text alignment. Users' understanding and debugging abilities are improved by this dual-pane structure, which makes it simple to compare the generated Python code with the original Arabic script.

\begin{figure}[h]
    \centering
        \centering
        \includegraphics[width=\textwidth]{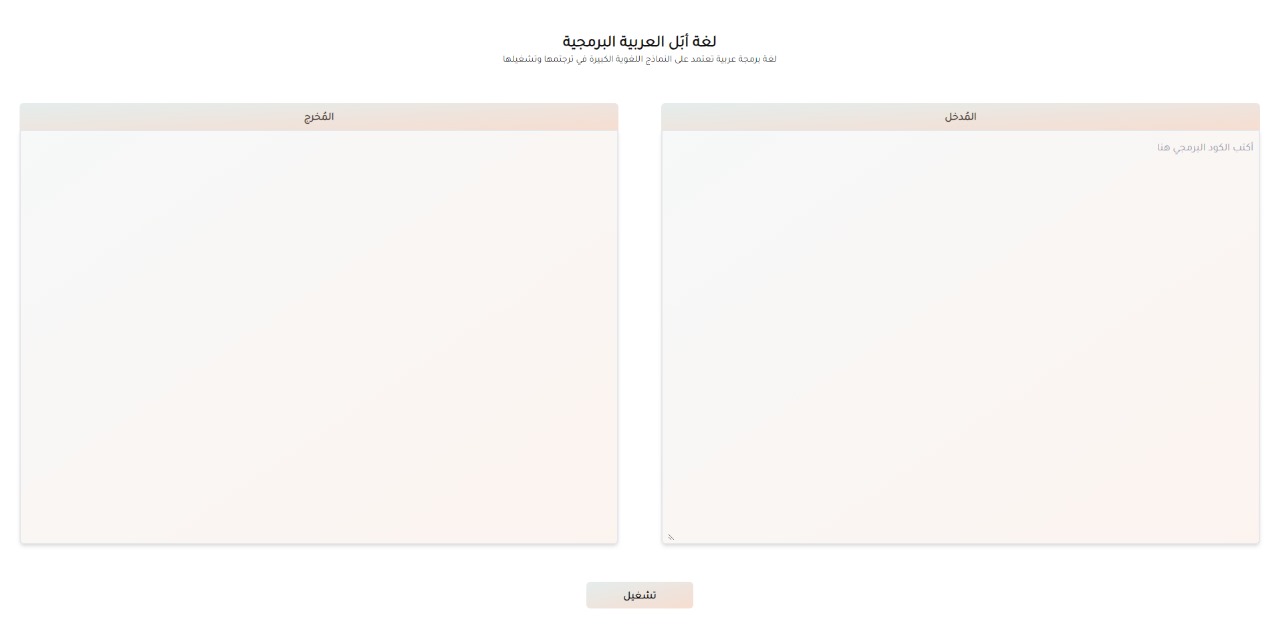}
        \caption{Interface of APL App}
        \label{fig:1}
\end{figure}

\begin{figure}[h]
    \centering
        \centering
        \includegraphics[width=\textwidth]{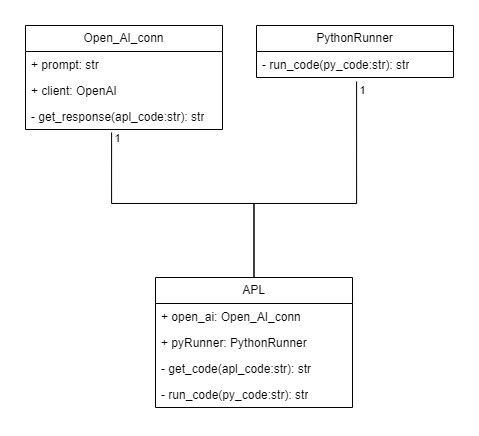}
        \caption{class diagram of the APL compiler}
        \label{fig:1}
\end{figure}

\section{Illustration}

In this section, we showcase the APL application of the LLM compiler in bridging linguistic barriers between Arabic and Python programming. The LLm compiler $APL$ has been tested under the progression from foundational programming concepts to more intricate operations illustrates not only the flexibility of the model but also its potential to democratize programming for Arabic-speaking individuals.

\begin{figure}[h]
    \centering
    \begin{subfigure}[b]{0.45\textwidth}
        \centering
        \includegraphics[width=\textwidth]{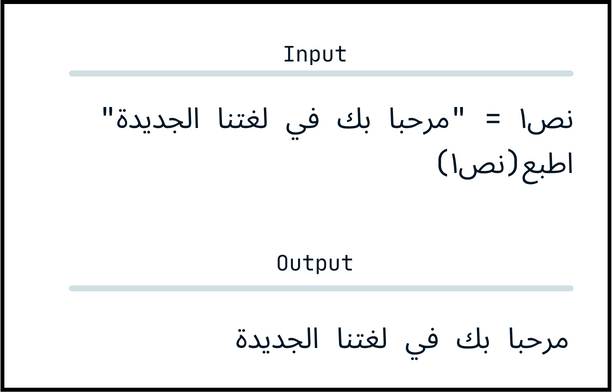}
        \caption{Example on print sentence }
        \label{fig:1}
    \end{subfigure}
    \hfill 
    \begin{subfigure}[b]{0.45\textwidth}
        \centering
        \includegraphics[width=\textwidth]{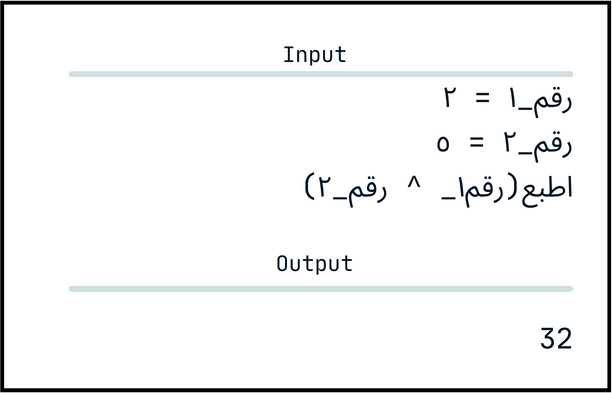}
        \caption{Example on simple arithmetic}
        \label{fig:2}
    \end{subfigure}
    \hfill 
    \begin{subfigure}[b]{0.45\textwidth}
        \centering
        \includegraphics[width=\textwidth]{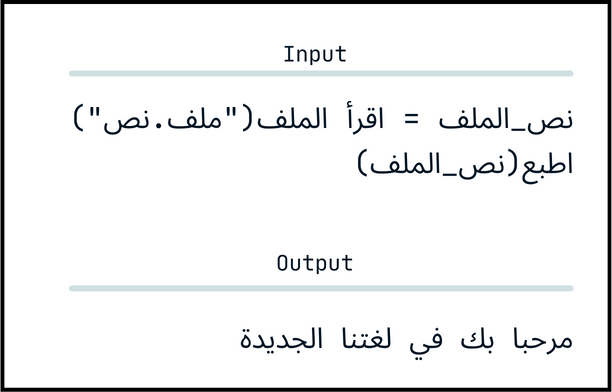}
        \caption{Example on reading a file}
        \label{fig:3}
    \end{subfigure}
    \hfill 
    \begin{subfigure}[b]{0.45\textwidth}
        \centering
        \includegraphics[width=\textwidth]{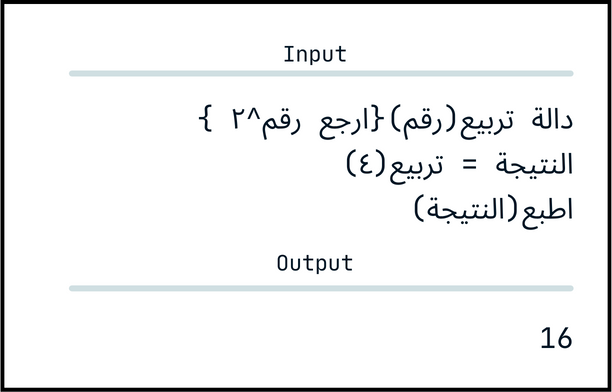}
        \caption{Example on making a function}
        \label{fig:4}
    \end{subfigure}
    \caption{Different code examples on input/output of the interface }
    \label{fig: simple examples}
\end{figure}

\begin{figure}[h]
    \centering
    \begin{subfigure}[b]{0.45\textwidth}
        \centering
        \includegraphics[width=\textwidth]{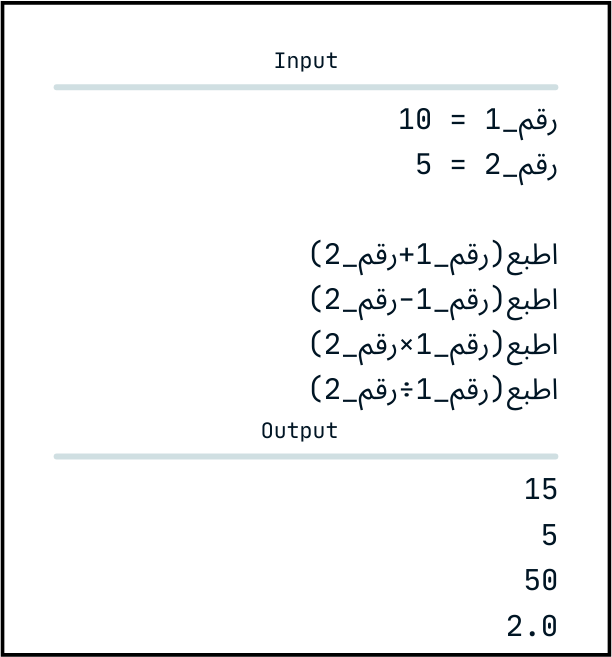}
        \caption{Example on making 2 vars then print them }
        \label{fig:1}
    \end{subfigure}
    \hfill 
    \begin{subfigure}[b]{0.45\textwidth}
        \centering
        \includegraphics[width=\textwidth]{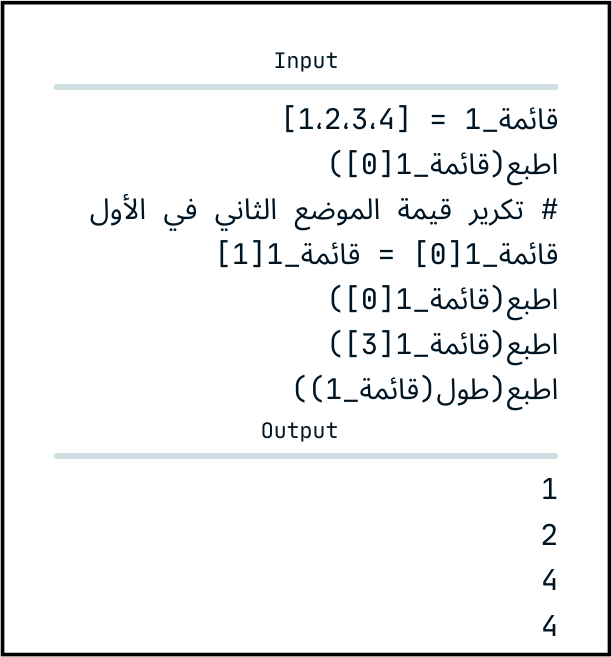}
        \caption{Example on making a list and change indexes and print them}
        \label{fig:2}
    \end{subfigure}
    \hfill 
    \begin{subfigure}[b]{0.45\textwidth}
        \centering
        \includegraphics[width=\textwidth]{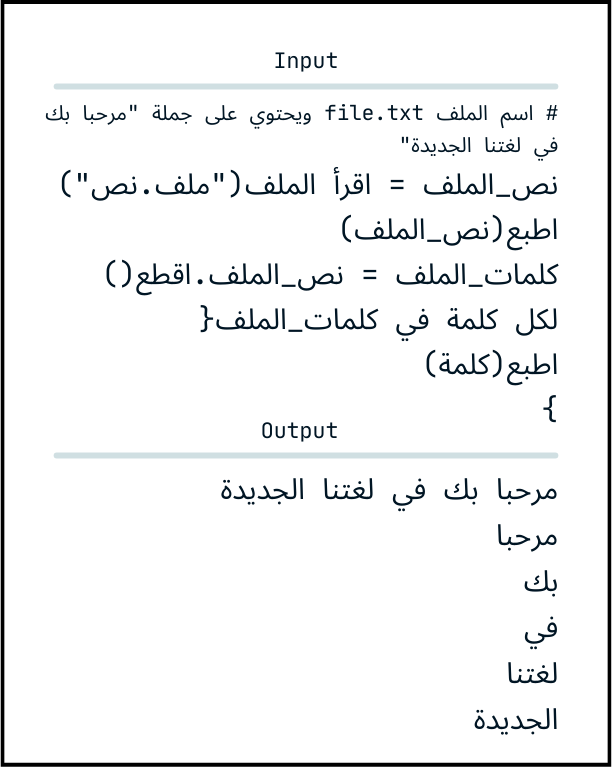}
        \caption{Example on reading a file and print the word by word}
        \label{fig:3}
    \end{subfigure}
    \hfill 
    \caption{Different complex code examples on input/output of the interface }
    \label{fig: diffcult examples}
\end{figure}

Figure\ref{fig: simple examples} demonstrates the foundational initiating with fundamental operations, such as file handling, arithmetic computations, string manipulations, and basic function definitions, this segment aims to solidify the groundwork. These examples emphasize the model's ability to comprehend and execute simple instructions, thereby laying a robust foundation for more complex tasks. Additionally, Figure \ref{fig: diffcult examples} shows more advanced Operations, building upon the basics, this segment delves into more sophisticated programming constructs. It includes the manipulation of data structures, control flow, and file operations with nuanced requirements. Demonstrating the model's adeptness at handling complexity

\section{challenges and future work}
The main challenges that faced us in the development of the compiler is the prompt engineering because it's facing Arabic text in two contexts the first is the programming language keywords and naming these need to be replaced with the English version so the python code could run without syntax errors, but on the other hand the Arabic text that was used in strings need to be kept as it is so if we print it or modify it in the code the Arabic characters still the same as the programmer added them. The second main problem we faced was the cost of using GPT-4 API where in the development phase this is not a big deal but in the deployment phase this will cost a of money to compile the APL code. In the future we could try to fine tune a open-source models so it reduce the costs of using paid APIs and also it will working on a specific task without the potential risk of getting useless responds from the non-generalized model.

\section{Conclusion}
In this paper we introduce APL an Arabic programming language that uses LLM as its compiler to translate the code into Python code. Also we produce a new API and interface to use the programming language.

\subsubsection*{Acknowledgments.} The heading should be treated as a
subsubsection heading and should not be assigned a number.

\end{document}